\documentclass[a4paper]{jpconf}
\usepackage{graphics,graphicx,dcolumn,bm,fleqn,epic,eepic,float,epsfig}
\usepackage{amssymb,amsmath,multirow,rotate,color,float}
\usepackage{epstopdf}
\usepackage{times}
\usepackage{color}
\usepackage{soul}                    
\definecolor{red}{rgb}{1,0,0}
\definecolor{green}{rgb}{0,1,0}
\definecolor{blue}{rgb}{0,0,1}

\begin{document}

\title{Reconstructing the intermittent dynamics of the torque in wind turbines}

\author{Pedro G.~Lind, Matthias W\"achter and Joachim Peinke}
\address{ForWind - Center for Wind Energy Research, Institute of Physics,
Carl-von-Ossietzky University of Oldenburg, DE-26111 Oldenburg, Germany}
\ead{pedro.lind@uni-oldenburg.de}



\begin{abstract}
We apply a framework introduced in the late nineties to analyze load 
measurements in off-shore wind energy converters (WEC).
The framework is borrowed from statistical physics and properly adapted
to the analysis of multivariate data comprising wind velocity, power 
production and torque measurements, taken at one single WEC.
In particular, we assume that wind statistics drives the fluctuations of 
the torque produced in the wind turbine and show how to extract an 
evolution equation of the Langevin type for the torque driven by the 
wind velocity.
It is known that the intermittent nature of the atmosphere, i.e. of
the wind field, is transferred to the power production of a wind energy 
converter and consequently to the shaft torque.
We show that the derived stochastic differential equation quantifies 
the dynamical coupling of the measured fluctuating properties as well
as it reproduces the intermittency observed in the data.
Finally, we discuss our approach in the light of turbine monitoring,
a particular important issue in off-shore wind farms.
\end{abstract}

\section{Introduction}

While wind energy can be taken as one of the best answers to the
world-wide energetic problem\cite{johnsonWindBook}, due to its particular 
physical features and turbulent nature it also presents challenging 
problems to be solved, even in more theoretical research fields such as 
physics and data analysis\cite{windenergyhandbook,davidmuecke}.
One of such open problems is the ability for developing methods that
reproduce the particular statistical properties shown in data series of 
power output or wind velocity measured at one wind turbine or wind energy 
converter (WEC).
As it is known\cite{windenergyhandbook}, since wind speed presents 
non-Gaussian fluctuations in time, the power output of one turbine
shows also this intermittent behavior\cite{patrickprl} making predictions
of energy production rather difficult.
Similarly, the intermittency of wind speed is also reflected
in the torque of the shaft.
Moreover, the loads applied by the wind on the WEC contribute 
significantly to determine the fatigue behavior and life expectancy 
of WECs\cite{ragan2007,moriarty08,freudenreich08}. 
Therefore, establishing good models for the intermittent 
evolution of the torque is an important task for better understanding 
and predicting the energy production and monitor the fatigue loads
in WECs.

In this paper we focus on the fluctuations of the torque, assuming
them as a direct result of atmospheric wind fields showing a high frequency
of extreme events (non Gaussian).
Recently, Milan et al\cite{muecke} have conjectured that the anomalous wind
statistics are responsible for the intermittent time evolution of the 
load, promoting additional fatigue of the turbine itself.
Here we show that this is indeed the case, by deriving an evolution 
equation for the series of torque measurements that is constrained to the
value of corresponding wind speeds.
To this end we use and analyze measurements of wind, power and torque
at one WEC of Alpha Ventus wind farm at the North Sea.

As we show below, initializing the differential evolution equation with 
the first values of our data series we are able to properly
reproduce the time series of the torque as well as its main statistical
features, including its intermittent behavior.
Our approach follows from the method proposed in 
Ref.~\cite{patrickprl,muecke} already applied to the power output of 
single WEC.

We start in Sec.~\ref{sec:data} by describing the data analyzed and
in Sec.~\ref{sec:method} the method is described in further detail.
Our results and comparative analysis is presented in Sec.~\ref{sec:results}
and conclusions and further discussions on this topic are given in
Sec.~\ref{sec:discuss}.
\begin{figure}[t]
\centering
\includegraphics[width=0.5\textwidth]{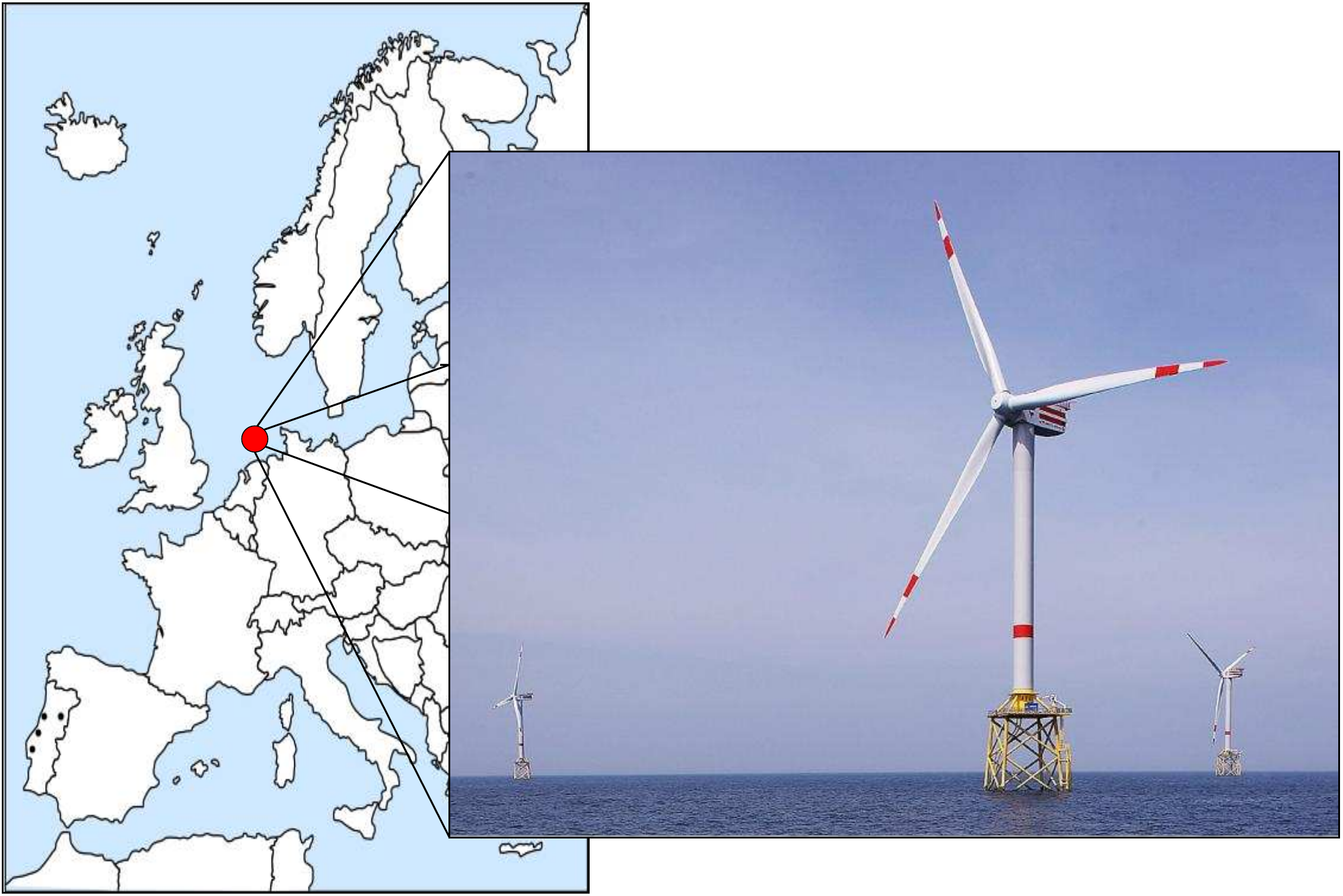}
\includegraphics[width=0.4\textwidth]{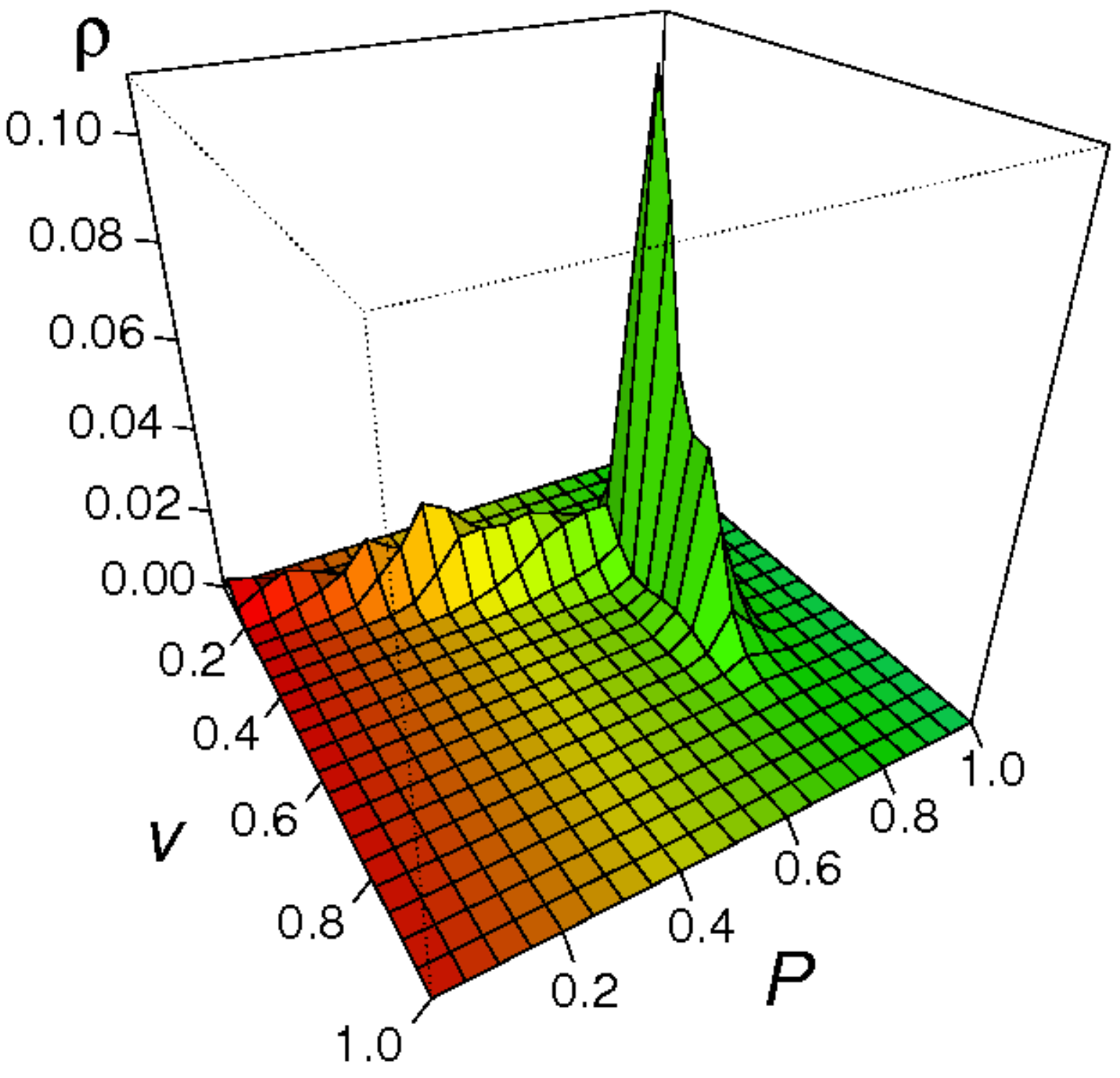}
\caption{\protect 
         (Left) Location of the R04 WEC at the Alpha Ventus wind
         farm (red bullet) [Photo: Sean Gallup/Getty Images].
         (Right)
         Joint probability density function as a function of the torque 
         and of the wind velocity. 
         All data was masked through normalization to the
         largest values (see text).}
\label{fig:Torque_1}
\end{figure}
\begin{figure}[htb]
\centering
\includegraphics[width=0.7\textwidth]{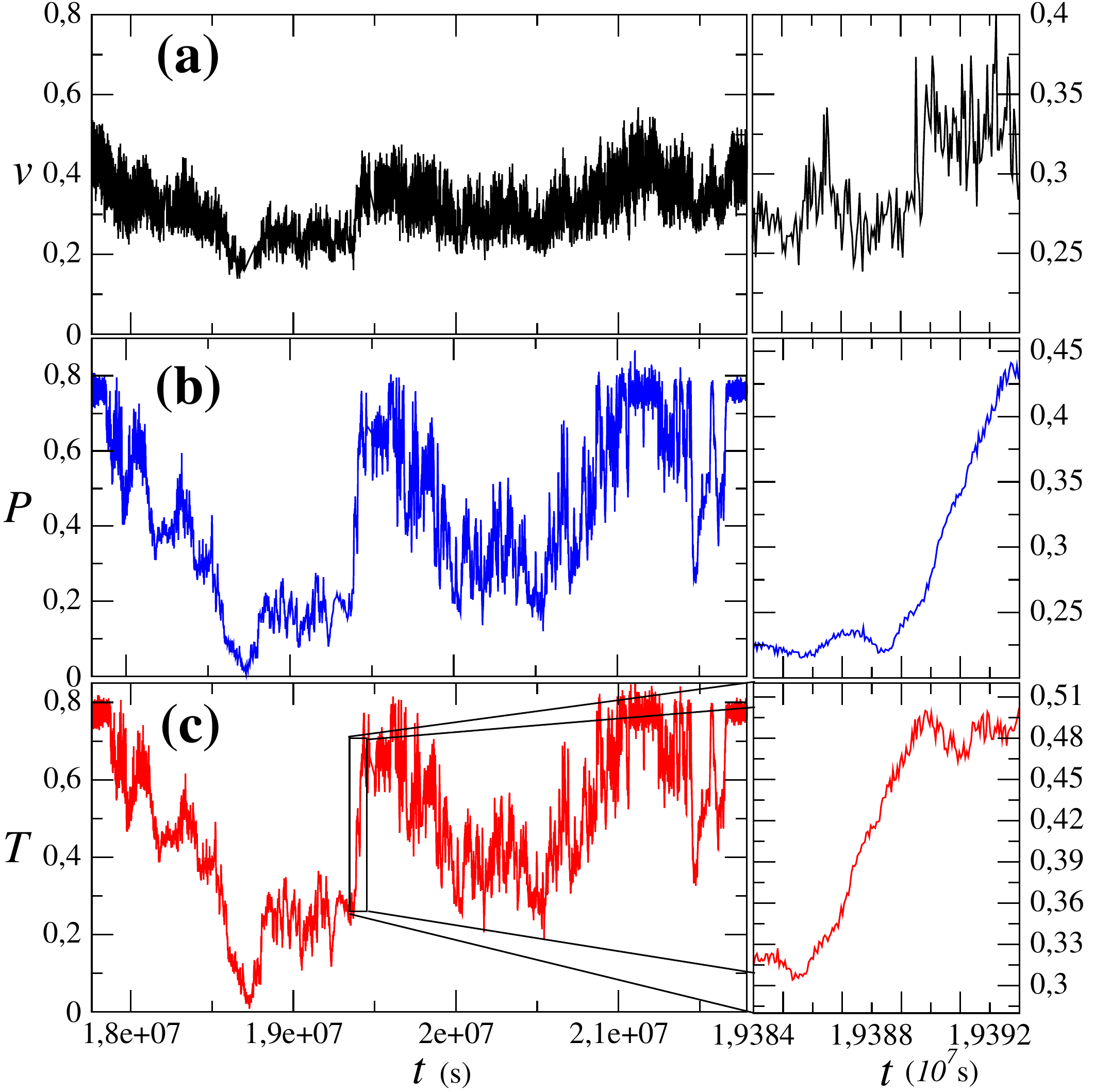}
\caption{\protect 
         Sketch of time series of 
         {\bf (a)} the wind velocity $v$,         
         {\bf (b)} the power output $P$ and
         {\bf (c)} the torque $T$.
         On the right a shorter time-interval of each series is 
         plotted to illustrated a stronger fluctuation of power and
         torque. All data was masked through normalization to the
         largest values of $v$, $P$ and $T$ respectively.}
\label{fig:Torque_2}
\end{figure}

\section{Data: the Alpha Ventus off-shore wind farm}
\label{sec:data}

The data analyzed in this paper comprehends three sets of measurements,
namely wind speed, power output and torque during the full month of
January 2013. The data was measured at one WEC of the Alpha Ventus wind 
farm (see Fig.~\ref{fig:Torque_1}, left), also known as Borkum West.
This wind farm is the first off-shore wind farm in Germany and it is 
located approximately at $54.3^o$N-$6.5^o$W. The torque is computed 
from the measurements of the power output $P$ and rotation number $n$, 
as $T=P/\omega$, with $\omega=n\pi/30$ the angular velocity of the 
operating shaft in units of rotations per minute.
The selected WEC was AV04 from Senvion, formely RePower.

The sampling rate of the power output and torque is $50$ Hz and the
sampling rate of the wind speed is $1$ Hz. 
Since we need to use the same sampling 
rate for all data series, we only consider power and torque 
measurements at instants for which a velocity measurement also exists
($1$ Hz).

All data series were analyzed according to all confidential protocols and
were properly masked through the normalization by their highest values.
Therefore the scientific conclusions are not affected by such data
protection requirements.

\begin{figure}[t]
\centering
\includegraphics[width=0.8\textwidth]{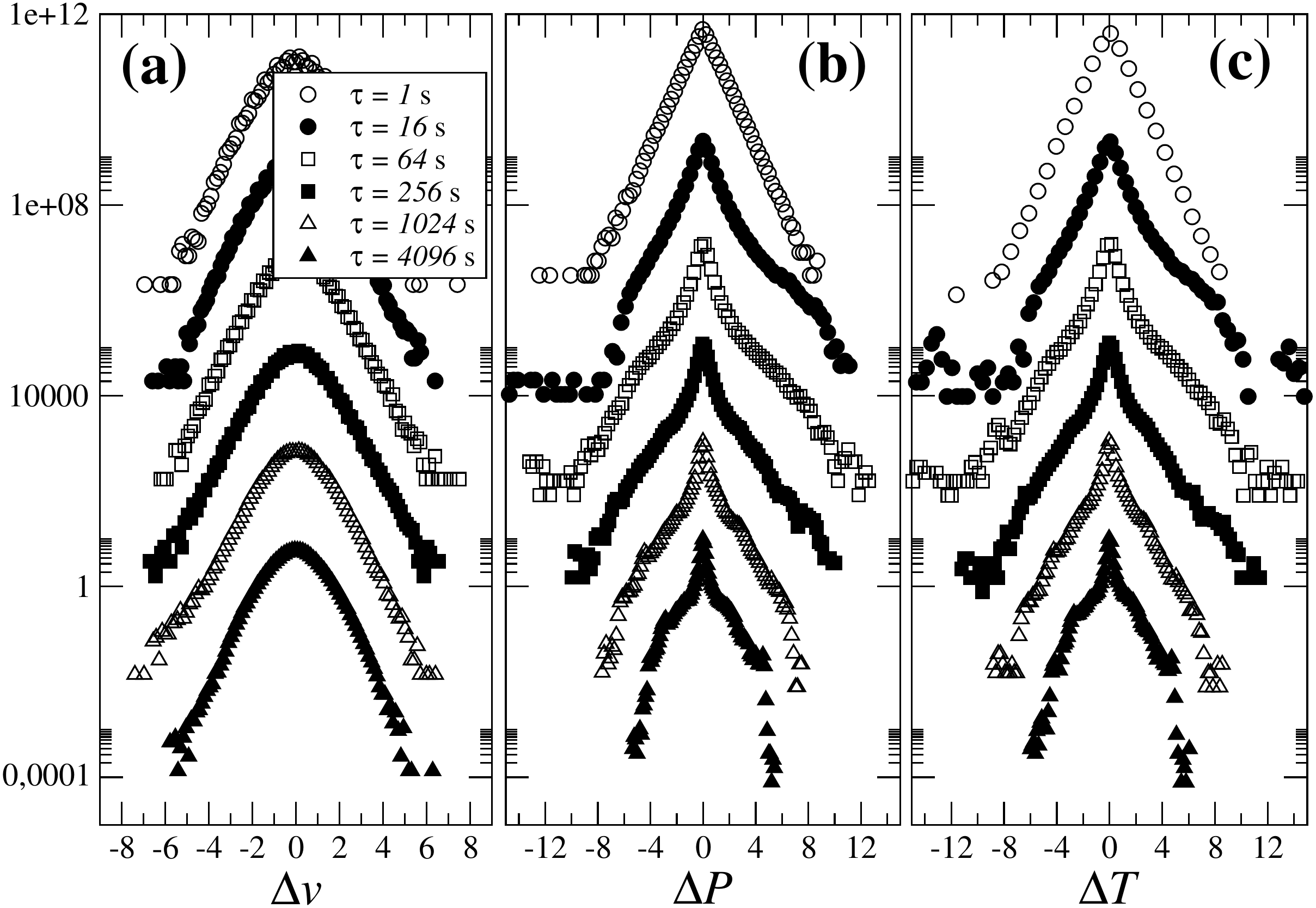}
\caption{\protect 
         Probability density functions (PDFs) for the increments of
         {\bf (a)} the wind velocity $\Delta v$,
         {\bf (b)} the power output $\Delta P$ and
         {\bf (c)} the torque $\Delta T$, for different time-lags 
         $\tau$.
         The increments are plotted in units of the
         corresponding standard deviation (see text).
         The shift in the vertical axis is for better
         visualization.}
\label{fig:Torque_3}
\end{figure}

The joint probability density function (PDF) $\rho(T,v)$ of both
the wind speed and torque measurements is shown in Fig.~\ref{fig:Torque_1}
(right) and is according to the torque-velocity curve known in
the literature\cite{windenergyhandbook,philipposter}. A time sampling
of each data series is plotted in Fig.~\ref{fig:Torque_2} (left) together
with an example of one time period where both torque and power 
change abruptly (right). 
These abrupt fluctuations are
the ones responsible for the intermittent behavior of the wind
energy production and wind loads which can be easily seen in the
increment statistics shown in Fig.~\ref{fig:Torque_3}.

To obtain the increment statistics of the torque $T$ we consider
torque difference taken within a fixed time-gap $\tau$, namely
\begin{equation}
\Delta T_{\tau}(t) = T(t+\tau)-T(t) ,
\label{increment}
\end{equation}
and similarly for the wind speed and power output.
As one sees from Fig.~\ref{fig:Torque_3}, for up to one hour or more,
the increment distributions are clearly non-Gaussian, particularly the
ones of torque and power. It is our purpose to provide a reconstruction
procedure that reproduces the same intermittent statistics at several 
time scales, in order to better quantify the torque fluctuations.
\begin{figure}[t]
\centering
\includegraphics[width=0.8\textwidth]{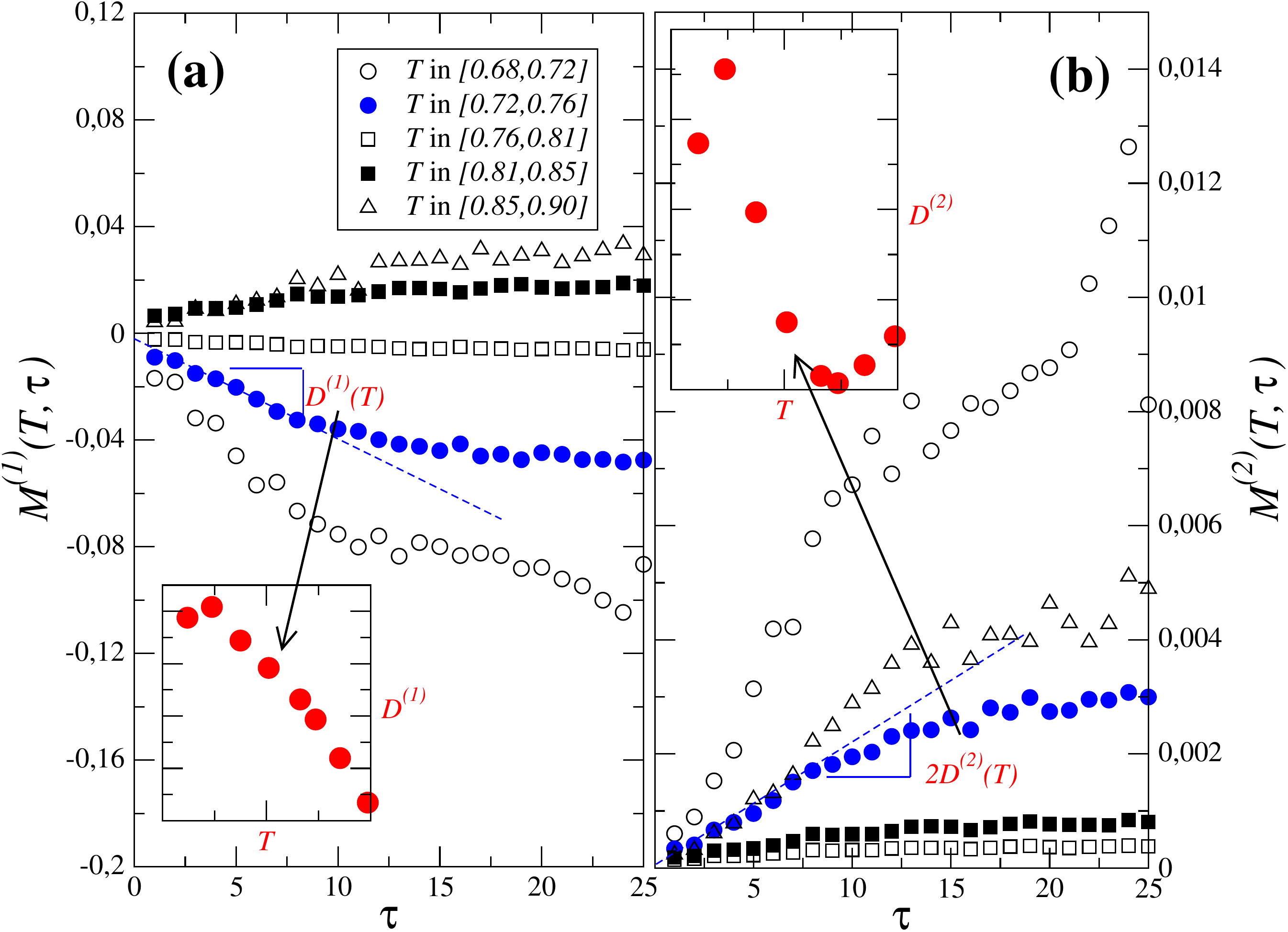}
\caption{\protect Conditional moments of the
         {\bf (a)} first and
         {\bf (b)} second order, for different values of the torque.
         In blue one illustrates the definition of the corresponding
         Kramers-Moyal coefficient, drift and diffusion, shown in
         the insets of (a) and (b) respectively.}
\label{fig:Torque_4}
\end{figure}

\section{Methodology: the conditioned Langevin approach}
\label{sec:method}

In 1997 a direct method to extract the evolution equation of stochastic
series of measurements was proposed by Peinke and Friedrich\cite{friedrich97}.
Since then several applications of this framework were proposed and
developed, ranging from turbulence modeling, to medical EEG monitoring
and stock markets. For a review in this methods see Ref.~\cite{physrepreview}
and references therein.
The method was also applied in the context of wind energy, where it 
was shown its ability to properly define the power characteristic of single
WECs\cite{muecke,anahua2008,raischel2013}.

The Langevin approach can briefly be described as follows.
Assume we have a set of measurements $X(t)$ in time $t$ of one 
particular property $x$ evolving according to the stochastic equation
\begin{equation}
\frac{dx}{d t}=D^{(1)}(x)+\sqrt{D^{(2)}(x)}\Gamma_t ,
\label{LangVect}
\end{equation}
where $\Gamma_t$ is a Gaussian $\delta$-correlated white noise,
i.e.~$\langle \Gamma(t)\rangle = 0$ and 
$\langle \Gamma(t)\Gamma(t')\rangle = 2\delta_{ij}\delta(t-t')$. 
Equation (\ref{LangVect}) is usually called a Langevin 
equation\cite{physrepreview}.
With such an {\it Ansatz}, one separates the deterministic contribution
to the evolution of $x$, given by the function $D^{(1)}$ (the drift), from
the stochastic fluctuations incorporated by function $D^{(2)}$, called
the diffusion.
The constant in $\delta$-correlation and the square root in 
the Langevin equation are usually chosen for convenience.

By simple integration of the Langevin equation, one easily extracts
a set of points similar to the sequence of, e.g., the torque measurements
in Fig.~\ref{fig:Torque_2}(c). But the problem here is the inverse one:
how can we arrive to a Langevin equation directly
from the analysis of the set of measurements $X(t)$?

The answer has two main steps. The first one concerns to test if there
is a time interval $t_{\ell}$ usually called the Markov length for which 
the succession of measurements are Markovian, i.e.~the next value only
depends on the present one and is independent of the values previous 
to it. Mathematically, to be Markovian means to fulfill the condition
\begin{equation}
\rho(X(t+t_{\ell})\vert X(t), X(t-t_{\ell}), X(t-2t_{\ell}), \dots) =   
\rho(X(t+t_{\ell})\vert X(t)) , 
\end{equation}
with $\rho$ representing the conditional probability density functions  
that can be extracted from histograms of the data set.
There are simple standard ways to perform this test\cite{physrepreview}.
When the measurements obey this Markov condition the next step
can be carried out. 
However, in the case the Markov test fails, for instance in the presence
of measurement noise\cite{boettcher2006}, the next step can still be 
applied, after taking some cautions that we do not mention here. 
See Ref.~\cite{lind2010} for details.

The second step, concerns the computation of both $D^{(1)}$ and $D^{(2)}$
that define Eq.~(\ref{LangVect}), done through the corresponding conditional
moments, illustrated in Fig.~\ref{fig:Torque_4}:
\begin{subequations}
\begin{eqnarray}
M^{(1)}(x,\tau) &=& \left\langle X(t+\tau)-X(t)  |_{X(t)=x} \right\rangle \\
M^{(2)}(x,\tau) &=& \left\langle (X(t+\tau)-X(t))^2  |_{X(t)=x} \right\rangle \\
\end{eqnarray}
\end{subequations}
where $\langle \cdot |_{X(t)=x} \rangle$ symbolizes a conditional averaging 
over the full measurement period.

Figure \ref{fig:Torque_4}a and \ref{fig:Torque_4}b shows the first
and second conditional moments respectively, for different values of the
torque, extracted from the data sets in Alpha Ventus.
As one sees, for the lowest range of values of $\tau$, the conditional 
moments depend linearly on $\tau$. Since the two functions in 
Eq.~(\ref{LangVect}) are, apart one multiplicative constant,
the derivative of the two corresponding
conditional moments with respect to $\tau$, namely
\begin{equation}
D^{(k)}(x)=\lim_{\tau\rightarrow0}\frac{1}{k!}\frac{M^{(k)}(x,\tau)}{\tau}
\label{DefCoefKM}\quad,
\end{equation}
with $k=1,2$ they can be directly extracted from the data sets.
As illustrated in Fig.~\ref{fig:Torque_4} with dashed lines, for each 
value $T$ both $D^{(1)}(T)$ and $D^{(2)}(T)$ are given by the slope of the 
linear interpolation of the corresponding conditional moments.
Important additional insight can be taken from such plots.
For instance, the linear fits of the conditional moments (dashed lines)
for the lowest range of $\tau$-values typically cross the zero-axis. 
The absence of an offset for the conditional moments gives evidence 
of the absence of measurement noise\cite{boettcher2006,lind2010}.

One important assumption however must be added: the set of measurements
must be stationary. This is of course {\it not} the case of power
and torque series.
To overwhelm this shortcoming, Milan et al propose to consider 
a Langevin equation, but restricted to a 
sufficiently confined range of wind velocities\cite{milanpriv}. 
Indeed, the statistical moments of the property being addressed
are approximately constant if only a narrow range of wind
velocities is considered.
Such variant leads to what we call the {\it conditioned} Langevin 
equation:
\begin{equation}
\frac{dT}{dt}=D^{(1)}(T,v)+\sqrt{D^{(2)}(T,v)}\Gamma_t,
\label{eq:langevinT}
\end{equation}
where, for our purpose, $T$ represents the torque on the WEC and 
$v$ is the wind velocity.

\section{Results: Reconstruction of the torque time series and statistics}
\label{sec:results}

Applying the methodology described in the previous section for
ranges of wind velocity  within $[\tilde{v},\tilde{v}+\Delta v]$
with $\Delta v=0.5$ and $\tilde{v}$ within the full range of
observed values, we derive the numerical estimates of the drift
$D^{(1)}$ and of the diffusion $D^{(2)}$ in Eq.~(\ref{eq:langevinT}).

\begin{figure}[t]
\centering
\includegraphics[width=0.45\textwidth]{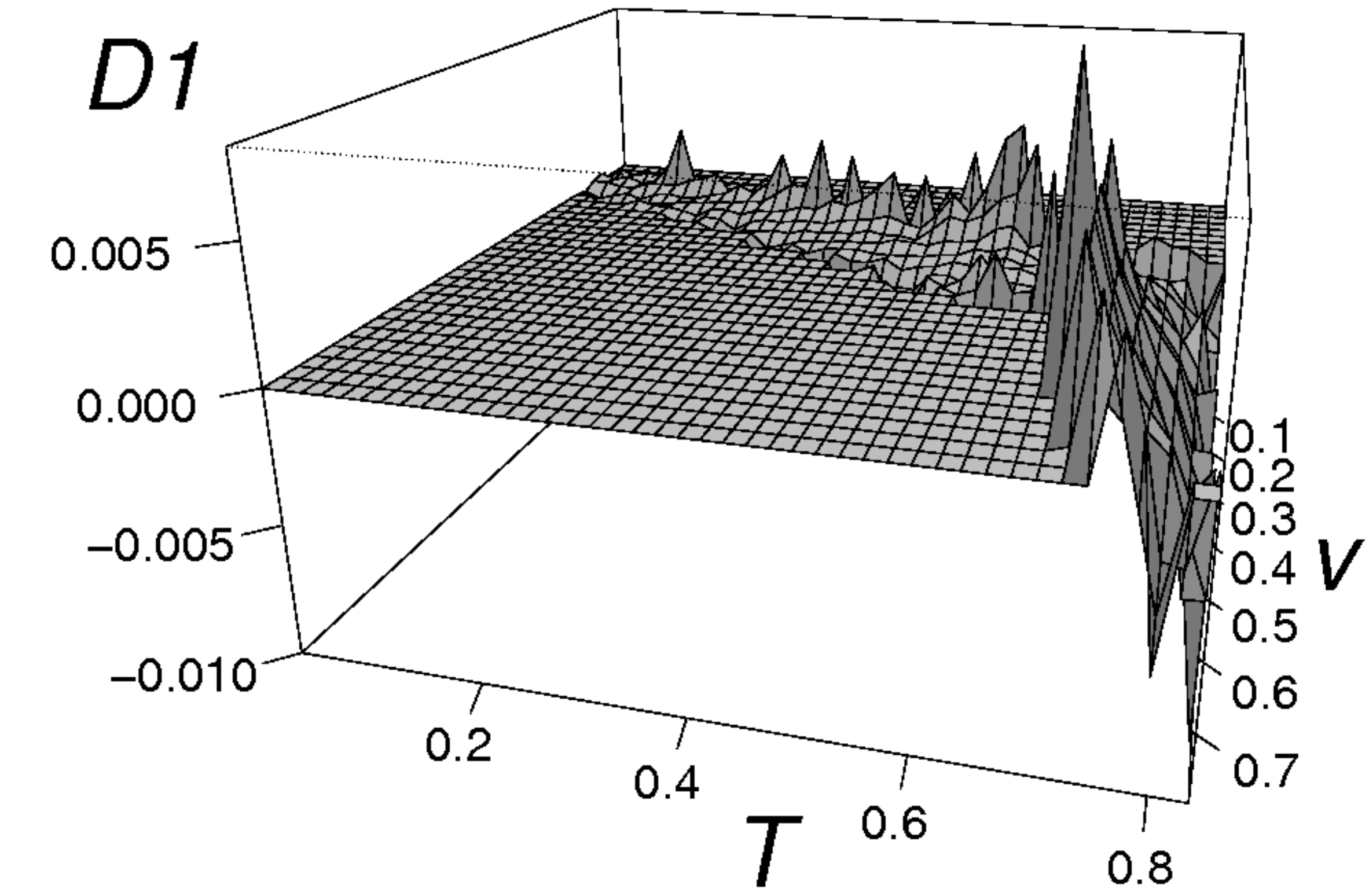}%
\includegraphics[width=0.45\textwidth]{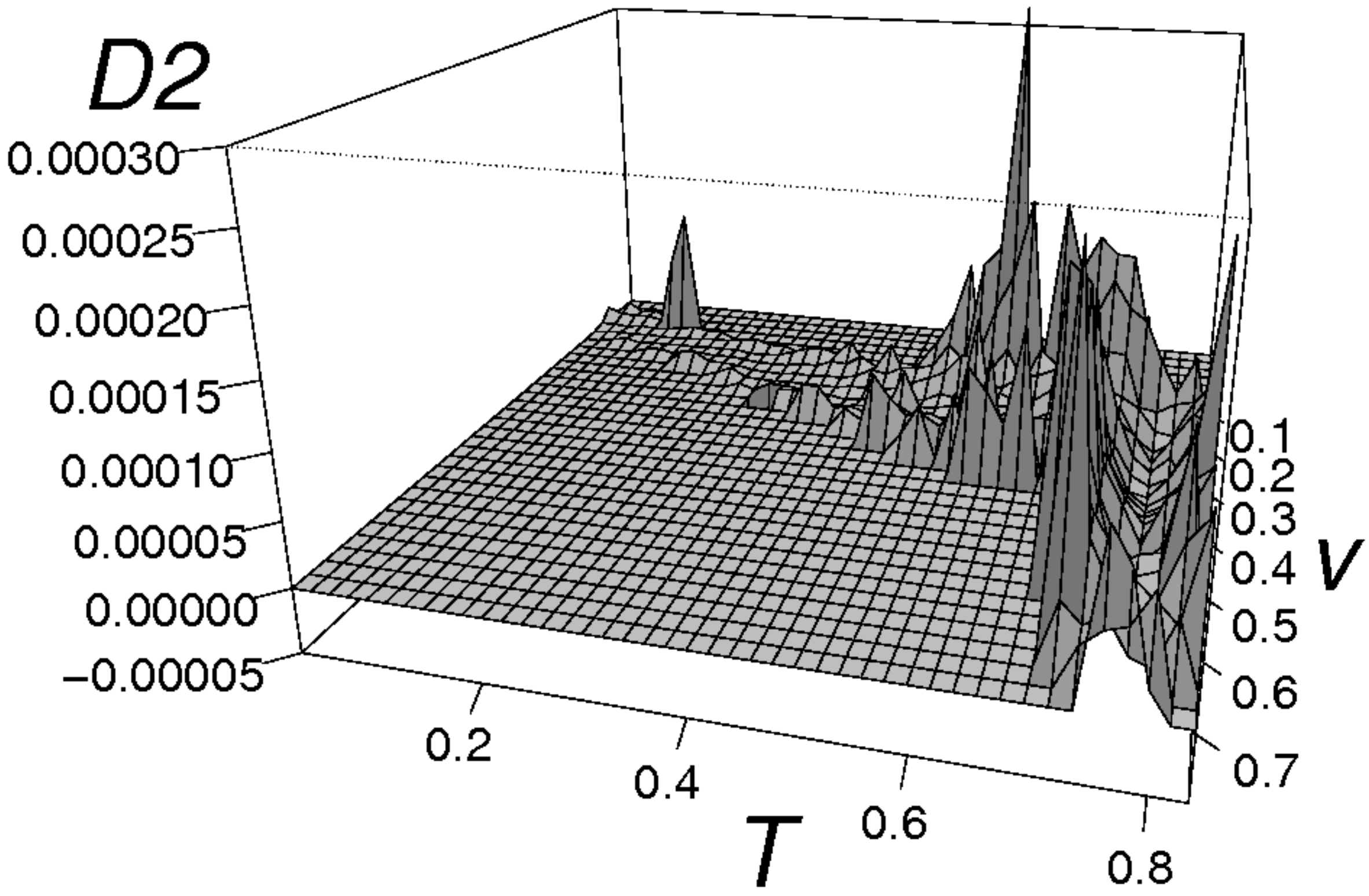}
\includegraphics[width=0.425\textwidth]{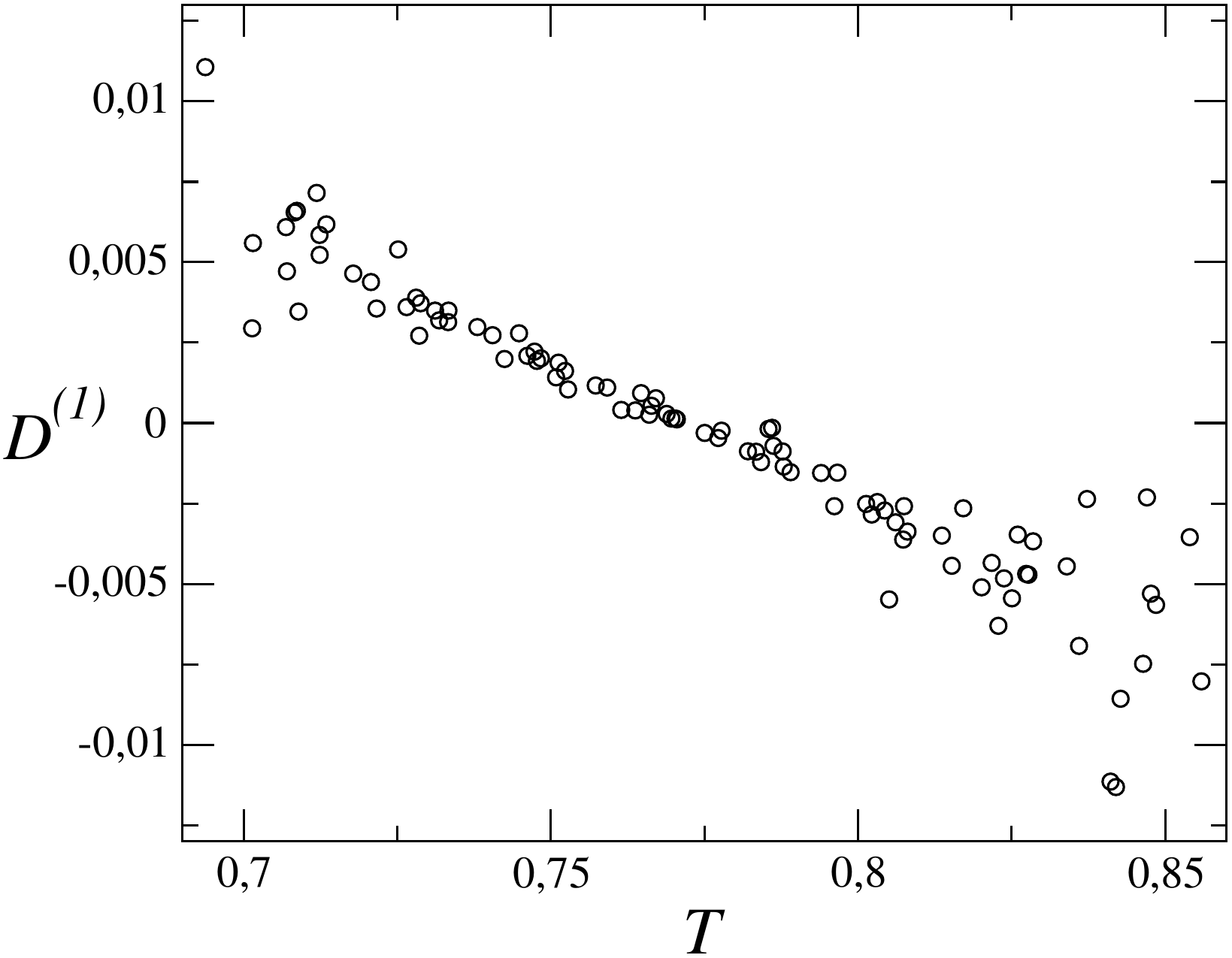}%
\hspace{0.5cm}%
\includegraphics[width=0.47\textwidth]{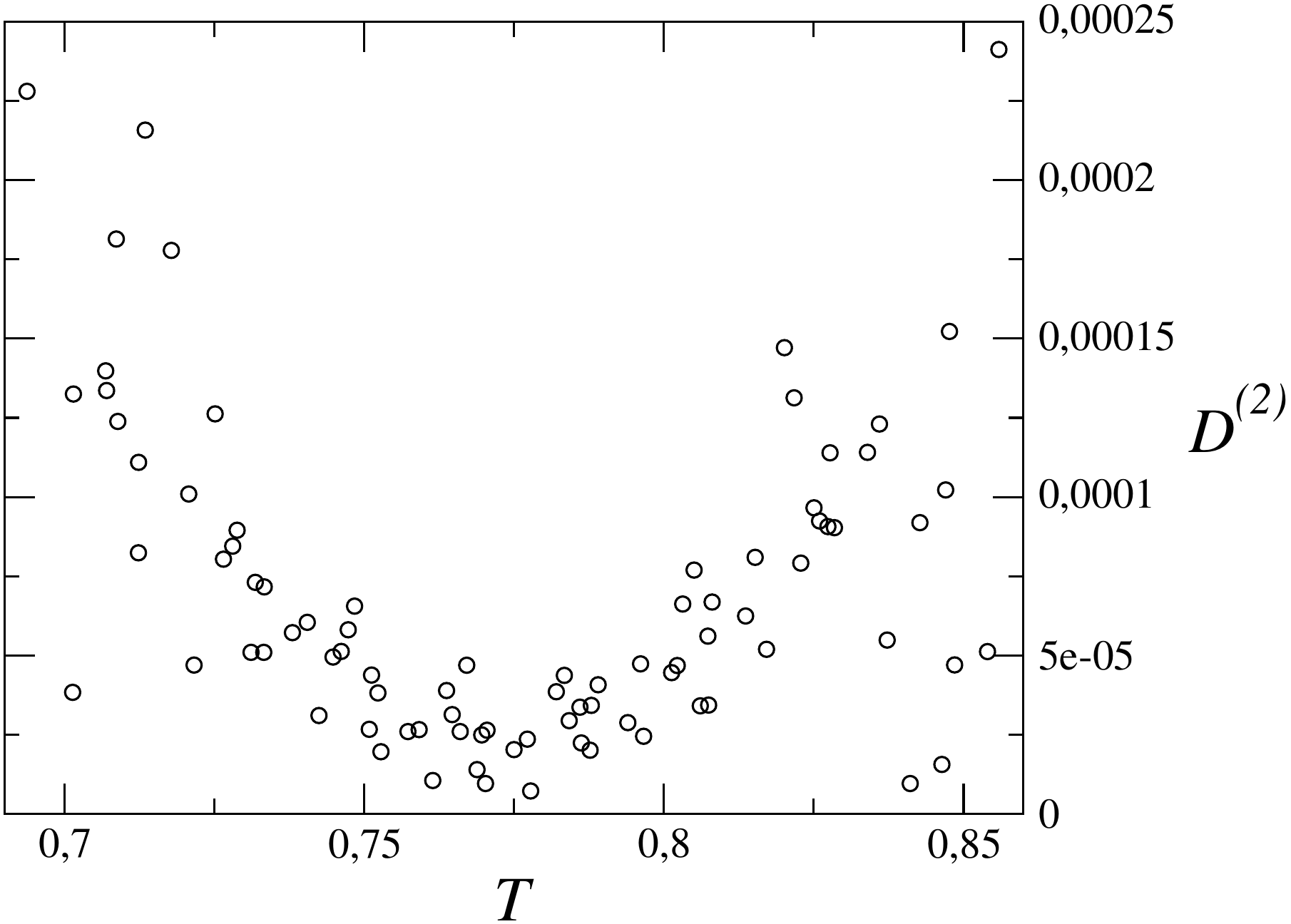}
\caption{\protect 
         (Top) Numerical result for the drift $D^{(1)}(T,v)$ and diffusion
         $D^{(2)}(T,v)$ in the Langevin equation from which the
         time series of the torque is reconstructed. 
         (Bottom) For the upper range of values of the wind velocity
         (around rated wind speed) one plots the projection
         of drift and diffusion on the $T$-axis. See text and
         Fig.~\ref{fig:Torque_6}.}
\label{fig:Torque_5}
\end{figure}

Figures \ref{fig:Torque_5} (top) show
the drift and diffusion coefficients respectively, each one
as a function of the wind velocity and of the torque.
While for low values of the velocity, $v\lesssim 0.3$,
both coefficients are poorly defined, due to the lack of 
sampling, in the most sampled range 
$0.3\lesssim v\lesssim 0.7$ (check with
Fig.~\ref{fig:Torque_1}) the drift and diffusion depend 
respectively linearly and quadratically on $T$. 
This dependence can be better seen in the two-dimensional
plots of Fig.~\ref{fig:Torque_5} (bottom).

Having extracted the functional dependence of both coefficients
$D^{(1)}$ and $D^{(2)}$ we are now able to describe the evolution
of the torque by keeping track of the wind velocity, simply through
an Euler-like discrete version of the conditioned Langevin equation.
We take the first measurement of both wind speed and torque as 
initial conditions
for the stochastic equation and integrate it with respect to $t$
using at each integration step the observed wind velocity.

\begin{figure}[t]
\centering
\includegraphics[width=0.85\textwidth]{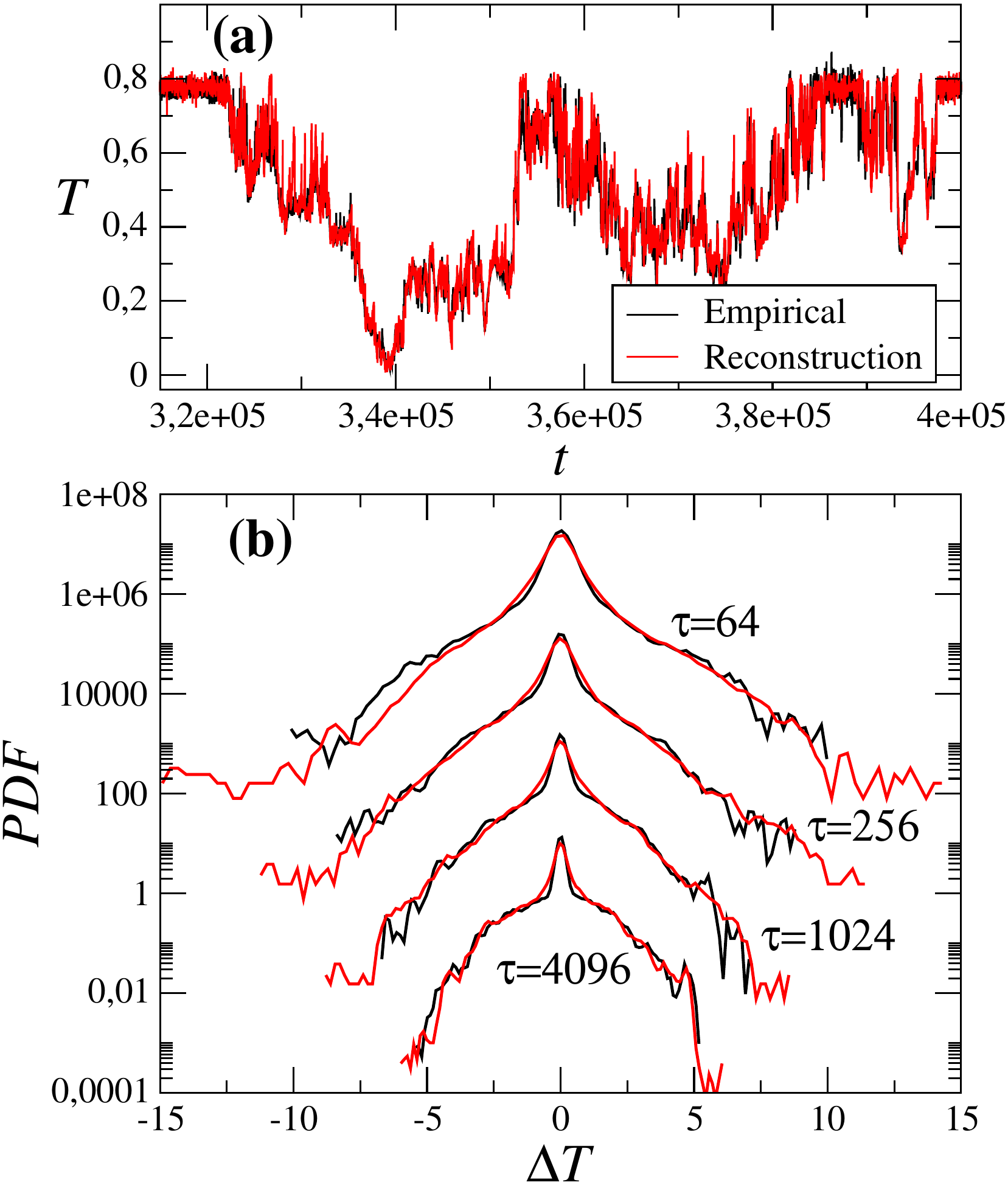}
\caption{\protect 
         Reconstruction of the torque (in red) of torque measurements (black):
         {\bf (a)} The explicit time-series of the torque and
         {\bf (b)} the probability density function of the increments 
         (fluctuations) of the torque within different time-lags.
         Time is in seconds, and the Torque increments is in units of
         corresponding standard deviations ($\sigma_{\tau}$).}
\label{fig:Torque_6}
\end{figure}

The reconstructed series are plotted in Fig.~\ref{fig:Torque_6}(a)
together with the empirical series of torque measurements.
Clearly, the reconstructed series are close to the real measurements.
Moreover, the statistical distribution of the increments
$\Delta T_{\tau}$ are also well reproduced for time
scales from seconds up to hours (Fig.~\ref{fig:Torque_6}(b)).
All in all, from Fig.~\ref{fig:Torque_6}, one can clearly
conclude the ability
for the conditioned Langevin model to properly describe
the evolution of the torque in one WEC.

\section{Discussion and conclusions}
\label{sec:discuss}

In this paper we show how to reconstruct the series
of torque measurements through a Langevin model conditioned
to the wind velocity. The model reproduces well
not only the series of torque measurements but also 
the intermittent feature of its increment statistics.
It should be noticed that the validity of Eq.~(\ref{eq:langevinT})
is not fully guaranteed, since Pawula condition\cite{risken}, 
$D^{(4)}=0$, was not tested.
This condition is necessary for assuming a Langevin
evolution equation. Still, even in the case Pawula theorem is
not fulfilled, the conditioned Langevin equation can be
taken as a first approximation of the stochastic evolution
of the torque in one WEC.

The reproduction of torque time series of WECs here described
is already significant, though the approximations used
for the drift and diffusion coefficients are of first order
only (see Fig.~\ref{fig:Torque_4}). Higher order approximations
are possible and would improve the reproduction 
further\cite{friedrich1,kantz,friedrich2,gottschall2008}. 
In these higher 
order corrections one considers the numerical values of 
the drift for the computation of the corresponding 
diffusion.

A critical remark should be stated at this point: while
the model in Eq.~(\ref{eq:langevinT}) properly reproduces the
stochastic evolution of the torque in WECs it depends on
the wind velocity measurements. On one hand, nacelle anemometer
wind velocity measurements are
typically much less accurate than the measurements of other
properties on the WEC, such as the pitch angle. On the other
hand, being always coupled to a measured property, in this
case the wind speed, the conditioned Langevin equation
is not able to provide straightforward forecasts
of loads even in the nearest time horizons.

Still, being able to properly describe the evolution of
the torque, one can use it for providing additional input
information for forecasting models, namely for training
neural networks constructed for torque forecast.

Another important next step from this study is to ascertain if
such an approach can be applied to other kinds of loads
in WEC, namely bending moments. These points can contribute
to improve monitoring protocols of WECs in off-shore wind
farms and will be addressed in the near future as the next 
steps.

\section*{Acknowledgments}

The authors thank Patrick Milan for useful discussions.
This work is funded by the German Environment Ministry as part of the
research project ``Probabilistic loads description, monitoring, and
reduction for the next generation offshore wind turbines (OWEA Loads)''
under grant number 0325577B.
Authors also thank Senvion for providing the data here analyzed.

\section*{References}

\bibliographystyle{iopart-num}
\bibliography{LindWaechterPeinke}

\end{document}